\documentstyle[12pt]{article}
\def\gev{{\rm\,GeV}}
\def\simgt{\rlap{\lower 3.5pt \hbox{$\mathchar\sim$}} \raise 1.5pt \hbox {$>$}}
\catcode`\@=11

\def\section{\@startsection{section}{1}{\z@}{3.5ex plus 1ex minus .2ex}
{2.3ex plus .2ex}{\large\bf}}

\def\thesection{\arabic{section}.}

\def\appendix{\setcounter{section}{0}
 \def\thesection{Appendix \Alph{section}:}
 \def\theequation{\Alph{section}.\arabic{equation}}}

\def\@citex[#1]#2{\if@filesw\immediate\write\@auxout{\string\citation{#2}}\fi
  \def\@citea{}\@cite{\@for\@citeb:=#2\do
    {\@citea\def\@citea{,\penalty\@m}\@ifundefined
       {b@\@citeb}{{\bf ?}\@warning
       {Citation `\@citeb' on page \thepage \space undefined}}%
\hbox{\csname b@\@citeb\endcsname}}}{#1}}

\def\citer{\@ifnextchar [{\@tempswatrue\@citexr}{\@tempswafalse\@citexr[]}}

\def\@citexr[#1]#2{\if@filesw\immediate\write\@auxout{\string\citation{#2}}\fi
  \def\@citea{}\@cite{\@for\@citeb:=#2\do
    {\@citea\def\@citea{--\penalty\@m}\@ifundefined
       {b@\@citeb}{{\bf ?}\@warning
       {Citation `\@citeb' on page \thepage \space undefined}}%
\hbox{\csname b@\@citeb\endcsname}}}{#1}}
\def\citenum#1{{\def\@cite##1##2{##1}\cite{#1}}}
\def\citea#1{\@cite{#1}{}}

\def\figitem{\@ifnextchar[{\@lfigitem}{\@figitem}}
\def\@lfigitem[#1]#2{\item{Fig.~\@figlabel{#1}.}\if@filesw
      { \def\protect##1{\string ##1\space}\immediate
        \write\@auxout{\string\figcite{#2}{#1}}\fi\ignorespaces}}
\def\@figitem#1{\item\if@filesw \immediate\write\@auxout
       {\string\figcite{#1}{\the\c@enumi}}\fi\ignorespaces}
\def\figcite#1#2{\global\@namedef{fb@#1}{#2}}
\let\citation\@gobble

\def\fcite{\@ifnextchar [{\@tempswatrue\@fcitex}{\@tempswafalse\@fcitex{[]}}}
\def\@fcitex#1#2{\if@filesw\immediate\write\@auxout{\string\citation{#2}}\fi
  \def\@fcitea{}\@fcite{\@for\@fciteb:=#2\do
    {\@fcitea\def\@fcitea{,}\@ifundefined
       {fb@\@fciteb}{{\bf ?}\@warning
       {Figure `\@fciteb' on page \thepage \space undefined}}%
\hbox{\csname fb@\@fciteb\endcsname}}}{#1}}
\let\figdata=\@gobble
\let\figliststyle=\@gobble
\def\figlist#1{\if@filesw\immediate\write\@auxout{\string\figdata{#1}}\fi
  \@input{\jobname.figs}}
\def\figliststyle#1{\if@filesw\immediate\write\@auxout
    {\string\figliststyle{#1}}\fi}

\def\@fcite#1#2{#1\if@tempswa , #2\fi}

\def\thefiglist#1{
 \par\section*{Figures \@mkboth{Figures}{Figures}}
\list
   {\m@th{Fig.\ \arabic{enumi}.\ \hfill}}
   {\settowidth\labelwidth{Fig.\ \m@th {#1}.\ }%
    \leftmargin\labelwidth
    \advance\leftmargin\labelsep
    \usecounter{enumi}}
    \def\newblock{\hskip .11em plus .33em minus -.07em}
    \sloppy
    \sfcode`\.=1000\relax}

\relax
\begin{document}
\thispagestyle{empty}
\hfill\vbox{\hbox{\bf TTP 95--02}
       \hbox{hep-ph/9505217}
       \hbox{April 1995}}
\vspace{1cm}
\begin{center}
{\large\bf SPECTRA OF BARYONS CONTAINING TWO HEAVY QUARKS  } \\
\vspace{1cm}
M.~L.~Stong \\
\vspace{1cm}
{\it Inst.~Theor.~Teilchenphysik, Univ. Karlsruhe, D-76128 Karlsruhe, Germany}
\\
\end{center}
\vspace{1cm}

\begin{center}
ABSTRACT \\
\end{center}

\vspace{0.5cm}
\begin{minipage}{0.05\textwidth}\end{minipage}
\begin{minipage}{0.90\textwidth}
{\small
The spectra of baryons containing two heavy quarks test the form of
the $QQ$ potential through the spin-averaged masses and hyperfine
splittings.  The mass splittings in these spectra are calculated
in a nonrelativistic potential model and the effects of varying the
potential studied.  The simple description in terms of light quark
and pointlike diquark is not yet valid for realistic heavy quark masses.}
\end{minipage}
\begin{minipage}{0.05\textwidth}\end{minipage} \\

\normalsize
\newpage

It is well-known that the quarkonium ($\Psi$ and $\Upsilon$) spectra
test the QCD potential between quark and antiquark.  In this case,
a phenomenological flavor-independent potential and a QCD-derived
hyperfine interaction can be used to describe the masses and splittings
of the $c \bar{c}$ and $b \bar{b}$ mesons \cite{mesons}.  Furthermore,
such models allow predictions to be made for the unobserved states in
the spectra \cite{quarko} and for the $B_c$ system \cite{bc}.
In the same way, the doubly-heavy baryons (those containing ccq or bbq)
test the potential between two quarks.
The one-gluon-exchange potential between heavy quarks differs from the
quark-antiquark potential by a relative color factor 2.
This relation does not hold at higher orders:  the one-loop corrections to
the $QQ$ and $Q\overline{Q}$ potentials are not equal \cite{gupta}.
The confining potential, which is determined phenomenologically,
should be studied for $QQ$ interactions as it has been for $Q \overline{Q}$.

\vspace{3mm}

Tests of the $QQ$ potential can be implemented not only in
the spin-averaged spectra of the doubly-heavy baryons, but
also by studying the hyperfine splittings of these particles.  This is
surprising, as the hyperfine interactions contain two types of terms:
those describing the spin-spin interaction between
the heavy quarks ($\sim 1/m_Q^2$), which depend on the $QQ$ potential
analogously to the quarkonium case, and those describing the
spin-spin interactions between heavy and light quarks ($\sim 1/m_Q m_q$),
which might be thought to depend on the light-heavy quark interaction
and not significantly on the heavy-heavy quark potential.
Nevertheless, the wave function of the light quark depends sufficiently on
the heavy quark pair to provide information on the $QQ$ binding.

\vspace{3mm}

Baryon spectroscopy with non-relativistic potential models
is relatively accurate, although not perfect.
The method is clearly limited by its use of a non-relativistic treatment
of the light quarks.  Nonetheless, such models provide a relatively good
description of the baryon spectra and stable predictions for the
hyperfine splittings $\Sigma_Q^* - \Sigma_Q$ and $\Sigma_Q - \Lambda_Q$ for
the baryons containing one heavy quark \cite{richard_AP}.
One potential choice which provides a good fit to the observed
baryon spectrum is the power-law form \cite{richard_potl}:
\begin{eqnarray} \label{eq:potl}
V(r_{ij}) & = & {1 \over 2} \sum_{i < j} \left( A + B r_{ij}^\beta
+ {C \over m_i m_j} \delta (r_{ij})
\vec\sigma_i \cdot \vec\sigma_j \right),
\end{eqnarray}
with
\begin{eqnarray} \label{eq:potl_params}\nonumber
\beta = 0.1, \hspace{1cm} A = -8.337 \gev, \hspace{1cm}
B = 6.9923 \gev^{1 + \beta}, \hspace{1cm} C = 2.572 \gev^2, \\ \nonumber
m_q = 0.3 \gev, \hspace{1cm} m_s = 0.6 \gev, \hspace{1cm}
m_c = 1.895 \gev, \hspace{1cm} m_b = 5.255 \gev.
\end{eqnarray}
The light and strange quark masses and the power $\beta$ are held
fixed at reasonable values, and the parameters $A$, $B$, and $C$
are obtained from a fit to the masses of the $N$, $\Delta$, and $\Omega^-$.
The charm and bottom quark masses are fit to the $\Lambda_c$ and
$\Lambda_b$.
This potential, when rescaled by the relative color factor of 2 between
$Q\overline{Q}$ and $QQ$ one-gluon exchange, fits
the $J/\Psi$ and $\Upsilon$ spin-averaged spectra as well.
Such power-law potentials are familiar from the description of $Q\overline{Q}$
bound states \cite{martin} and give the choice of $\beta$ used above.
The value of $C$ given above is however significantly larger than
that obtained from a fit to the $J/\Psi$--$\eta_c$ hyperfine splitting, which
is 1.172 \cite{richard_potl}.

\vspace{3mm}

Although this potential fits the known baryon spectrum, it is not
convenient for study of the effects on the spectra of variation of
the $QQ$ potential.  In particular, it is interesting to be able to
vary separately the Coulomb-like small distance contribution to the
potential and the long-distance confining term.
The potential between light and heavy quarks is therefore left as in
Eq.~(\ref{eq:potl}), while the $QQ$ interaction is replaced with the
Cornell form \cite{cornellp}:
\begin{eqnarray} \label{eq:cornell}
V_{QQ}(r) & = & {1 \over 2} \left( - {k \over r} + a r + c
+ {C \over m_Q^2 } \delta (r) \vec\sigma_1 \cdot \vec\sigma_2 \right),
\\ \nonumber
a = 0.2005 \gev^2, \hspace{1cm} k & = & 0.4274, \hspace{1cm} c = -1.0455 \gev .
\end{eqnarray}
The parameters $a$, $k$, and $c$ are obtained from a fit to the $J/\Psi$ and
$\Upsilon$ spectra, with the quark masses and $C$ held fixed to the values
given above.
The effects on the hyperfine splittings of small variations in $a$ and $k$
are then studied.  These variations, as well as the power-law type potential
given above, are plotted in Figure~\fcite{fig:potl} for comparison.

\vspace{3mm}

In the limit of heavy quark mass, the heavy $QQ$ pair is more easily
excited to higher-energy states than the light quark.
In the harmonic oscillator potential, for example, the ratio of
the excitation energies for light and heavy degrees of
freedom is $\sqrt{(2m_h + m_l)/3 m_l}$ \cite{richard_3bp}.
The states considered here are those containing two identical heavy
quarks ($cc$ and $bb$ rather than $bc$) in a relative S-wave state,
so that Fermi statistics require that the heavy quarks be in a spin-1
(symmetric) state.  In the standard notation the lowest-energy
states for $QQu$ or $QQd$ are $\Xi_{QQ}$ ($\Xi^*_{QQ}$) for spin 1/2 (3/2),
and for $QQs$ they are denoted by  $\Omega_{QQ}$ ($\Omega^*_{QQ}$ ).
The heavy-heavy hyperfine terms cancel in the difference between spin-3/2
and 1/2 states as the expectation value of the spin-spin operator,
$<\vec\sigma_1 \cdot \vec\sigma_2>$, is identical for these two cases.
The hyperfine splittings discussed below thus originate solely from the
light-heavy quark interactions.
Three different mass splittings are
considered here for each combination of heavy and light quark flavors.
The first is the difference between the centers-of-mass of the states
in which the $QQ$ pair is in 2S and 1S states.  The other two are
the hyperfine splittings between the spin-3/2 and spin-1/2 baryons for
the 2S and 1S $QQ$ configurations.

\vspace{3mm}

Once the potentials are set, the solution of the three-body problem
can be approached using a variety of numerical methods \cite{richard_3bp}.
The Born-Oppenheimer method is an efficient solution method
for the case of two heavy and one light quark.
The wavefunction is split into heavy- and light-quark degrees of freedom,
\begin{eqnarray} \label{eq:bo_split}
\Psi(\rho , \lambda ) = \sum_n \phi_n(\rho ) f_n(\rho , \lambda ),
\end{eqnarray}
where $\rho$ is the distance between the two heavy quarks ($m_1$ and $m_2$),
and $\vec{\lambda} = (\vec{r}_3 - \vec{R}_{12})
\sqrt{ m_3(m_1\!+\!m_2)/\mu_{12}(m_1\!+\!m_2\!+\!m_3) }$
is proportional to the distance between the light quark $m_3$
and the center-of-mass $\vec{R}_{12}$ of the heavy-quark pair.
This choice of variables separates the non-relativistic kinetic
energy terms:  $p_1^2/(2 m_1) + p_2^2/(2 m_2) + p_3^2/(2 m_3)
= [ p_\rho^2 + p_\lambda^2 ]/(2 \mu_{12})$, where $\mu_{12}$ is
the reduced mass of the two heavy quarks.
The dependence on the light-quark mass has been absorbed into the
definition of $\lambda$.
The light quark wavefunction $f$ is found for fixed $\rho$
\begin{eqnarray} \label{eq:BO-lq}
\left[ {1 \over 2 \mu_{12}} p_\lambda^2 + V(r_{13}) + V(r_{23}) \right]
f_n(\rho , \lambda ) & = & \epsilon_n (\rho) f_n(\rho , \lambda ),
\end{eqnarray}
where $r_{i3}$ is the distance between the light quark and heavy quark $i$.
The energy of the $QQq$ state is approximated by:
\begin{eqnarray} \label{eq:BO-hq}
\left[ {1 \over 2 \mu_{12}} p_\rho^2 + V_{QQ}(\rho) + \epsilon_0 (\rho)
+ < f_0 | {1 \over 2 \mu_{12}} p_\rho^2 | f_0 > \right] \phi_0 (\rho ) =
E \phi_0 ( \rho ).
\end{eqnarray}
Keeping only the zeroth term in the expansion of $\Psi$ corresponds to
the so-called ``uncoupled adiabatic'' \cite{bo-ua} approximation.

\vspace{3mm}

The Born-Oppenheimer method integrates over the distances between
the heavy quarks in the second step (Eq.~(\ref{eq:BO-hq})), so that
the above approximation differs from that in which the heavy-quark
pair behaves as a spin-one pointlike diquark, as it would for
infinitely heavy quark mass.
Because this method does not depend on the limit of infinite $m_Q$, it
works well already for the charmed baryons.
In the limit $m_{1,2} \gg m_3$ these equations approach the expected form:
The distances $r_{i3}$ in Eq.~(\ref{eq:BO-lq}) are independent of the
quark masses, and the term $p_\lambda^2/2\mu_{12} \rightarrow p_3^2/2m_3$,
so that the light-quark wavefunction is independent of the distance
$\rho$ between the heavy quarks.
Eq.~(\ref{eq:BO-hq}) then reduces to the usual non-relativistic
two-body equation.
In the limit of infinite $m_Q$, the $QQq$ spectra can be simply related to the
meson spectra \cite{sav_wise}.  The splittings obtained in this way are
however significantly smaller than those obtained from the potential
model calculation, and the masses of the spin-averaged states somewhat higher.
This limit gives the values:
$M_{\Xi_{cc}^*} = 3.811 \gev$, $M_{\Xi_{cc}} = 3.742 \gev$,
$M_{\Xi_{bb}^*} = 10.484 \gev$, and $M_{\Xi_{bb}} = 10.460 \gev$
\cite{hqet_spec}.

\vspace{3mm}

The Coulomb and linear terms in the $QQ$ potential are now varied by
introducing parameters $x_a$ and $x_k$:
\begin{eqnarray} \label{eq:cornell_var}
V_{QQ}(r) & = & {1 \over 2} \left( - x_k {k \over r} + x_a a r + c
+ {C \over m_Q^2 } \delta (r) \vec\sigma_1 \cdot \vec\sigma_2 \right),
\end{eqnarray}
which are varied between 0.8 and 1.2 to determine the effect on the mass
differences M(2S)--M(1S), $\Delta$M(1S), and $\Delta$M(2S).
Variations in $c$ do not affect the splittings, only the masses
themselves.  Variations in $C$ obviously cause proportional changes
in the hyperfine splittings.  The effect of $C$ on the 2S--1S mass
difference comes from the $1/m_Q^2$ hyperfine terms only, and is small.
These results of the $x_a$ and $x_k$ variations are presented in
Table~\ref{tab:var}, along with the percent change
$100(\Delta {\rm M}(x_i\!=\!1.2)-\Delta {\rm M}(x_i\!=\!0.8))/
\Delta {\rm M}(x_i\!=\!1)$.
The values for $x_a=x_k=1$ are in
agreement with previous calculations \cite{prev_calc} when the power-law
type potential is used for the $QQ$ interaction.

\vspace{3mm}

The hyperfine splittings are analogous to those in a heavy-light
meson system, so that in the limit of infinitely heavy $m_Q$ they
should approach $\delta/m_Q$, where $\delta$ is independent of heavy-quark
flavor.  The mass splittings in the table give
$\delta$(ccq,1S) = 0.242 and $\delta$(bbq,1S) = 0.305.
These states are not yet
heavy enough to be accurately described by the simple heavy-quark limit.
The values of $\delta$ extracted for these and for heavier masses are
shown in Fig.~\fcite{fig:masslim}, where $\delta$ may be seen to approach
the heavy-mass limit for $m_Q \simgt$ 15--20 GeV.
This is in contrast to what is known from meson spectroscopy, where the
simple $1/m_Q$ description of the splittings works even for
the strange quark.  Although the masses of the diquarks are
quite large ($\sim 2m_Q$),
the radii for these states are also large for
realistic quark masses, so that the simple application of the
infinite-mass limit is not yet valid here.
Typical radii $<\!r\!>$ for $QQ$ states are 2.4 GeV$^{-1}$
for $cc$ 1S, and 1.5 GeV$^{-1}$ for $bb$ 1S.
For $Q \overline{Q}$ states, typical radii are 1.7 GeV$^{-1}$
for $c \bar{c}$, and 1.0 GeV$^{-1}$ for $b \bar{b}$.

\vspace{3mm}

The percent changes in the mass splittings are given in Fig.~\fcite{fig:var}
for $QQq$.  The sensitivity of the $QQs$ states to changes
in the potential is almost exactly that of the $QQq$ states.
M(2S)--M(1S) is more sensitive to changes in the $QQ$ potential than
either of the hyperfine splittings.  This mass difference is essentially
the mass splitting between the 1S and 2S states of the two-body $QQ$
potential, so that the strong dependence here is unsurprising.
The dependence of M(2S)--M(1S) on $x_k$ is about one-quarter(half) that
on $x_a$ for the cc (bb) states.  The bb pair is more closely bound
than the cc so that these states depend more strongly on the Coulomb
($x_k$) part of the potential.

\vspace{3mm}

The hyperfine splitting in the 1S states, $\Delta$M(1S),
depends more strongly on $x_a$ than on $x_k$ for the cc case,
and on $x_k$ than on $x_a$ for bb.  The strong dependence of the bb
1S splitting on the Coulomb term in the potential reflects the
closer binding of the b quarks.
On the other hand, $\Delta$M(2S) depends in both cases more
strongly on $x_a$ than on $x_k$.  This is a reflection of the larger
radius ($<r> \sim $3.0 GeV$^{-1}$) of the $bb$ 2S state.

\vspace{3mm}

In conclusion, the spectra of doubly-heavy baryons have been studied in
a non-relativistic potential model, with emphasis on the sensitivity of
the mass splittings of these states to variations in the interaction
between the two heavy quarks.  Although the hyperfine splittings are
between the two possible spin states of the light quark relative to
a heavy spin-1 system, it was found that dependence on the $QQ$ interaction
remains.  This dependence is however rather small.  The 2S--1S mass
difference is more sensitive to variations in the potential than the
hyperfine splittings.
A comparison of the splittings between the center-of-mass of
the 2S and 1S states and the hyperfine splittings in these levels
exhibits different dependences on the strengths of the separate parts
of the potential (Coulomb vs.\ confining terms).
More importantly, it is demonstrated that the use of the infinitely
heavy $m_Q$ limit, where the heavy quarks form a pointlike diquark
and the baryon as a whole a meson-like system, is not yet
valid for realistic heavy quark masses.

\vspace{1cm}
{\noindent\large\bf Acknowledgements}
\vspace{3mm}

\noindent
I would like to thank Marek Je\.zabek for suggesting the question
of the sensitivity of the hyperfine splittings to the $QQ$ potential.

\begin{thefiglist}{99}
\figitem{fig:potl} The $QQ$ potentials used in this paper.
Solid lines denote the power-law (Eq.~(\ref{eq:potl})) and
Coulomb-plus-linear potentials (Eq.~(\ref{eq:cornell}))
without any variation.  The dot-dashed (dotted) lines show the effect of
the variation of $x_k$ ($x_a$) between 0.8 and 1.2.
\figitem{fig:masslim}  The quantity $\delta = m_Q \Delta$M(1S)
as a function of heavy-quark mass.
\figitem{fig:var}  The percent changes in the mass splittings
$\delta {\cal M}(x_i) = 100 ({\cal M}(x_i)/{\cal M}(1) - 1)$.
The labels 1, 2, 3 refer to M(2S)--M(1S), $\Delta$M(1S), and
$\Delta$M(2S), respectively, while a and k indicate variations in $x_a$
or $x_k$.
\end{thefiglist}

\begin{table}[h]
\begin{tabular}{|r|r|r|r|r|} \hline

    & $x_a$  & M(2S)-M(1S) & $\Delta$M(1S) & $\Delta$M(2S) \\ \hline \hline
ccq &  0.80  & 0.394       & 0.125          & 0.0802  \\ \hline
    &  1.00  & 0.431       & 0.128          & 0.0832  \\ \hline
    &  1.20  & 0.467       & 0.130          & 0.0861  \\ \hline
 & \% change & 16.9        & 4.2            &  7.0    \\ \hline \hline
bbq &  0.80  & 0.280       & 0.0571         & 0.0417  \\ \hline
    &  1.00  & 0.304       & 0.0581         & 0.0430  \\ \hline
    &  1.20  & 0.328       & 0.0590         & 0.0441  \\ \hline
 & \% change & 15.7        & 3.3            & 5.6     \\ \hline \hline
    & $x_k$  &             &                &         \\ \hline
ccq &  0.80  & 0.423       & 0.126          & 0.0828  \\ \hline
    &  1.00  & 0.431       & 0.128          & 0.0832  \\ \hline
    &  1.20  & 0.440       & 0.129          & 0.0837  \\ \hline
 & \% change & 3.8         & 2.3            & 1.1     \\ \hline \hline
bbq &  0.80  & 0.294       & 0.0565         & 0.0425  \\ \hline
    &  1.00  & 0.304       & 0.0581         & 0.0430  \\ \hline
    &  1.20  & 0.316       & 0.0593         & 0.0435  \\ \hline
 & \% change & 7.4         & 4.8            & 2.4     \\ \hline \hline
\end{tabular}
\caption{Mass splittings (GeV) in the $ccq$ and $bbq$ systems as a function of
the parameters $x_a$ and $x_k$.}
\label{tab:var}
\end{table}

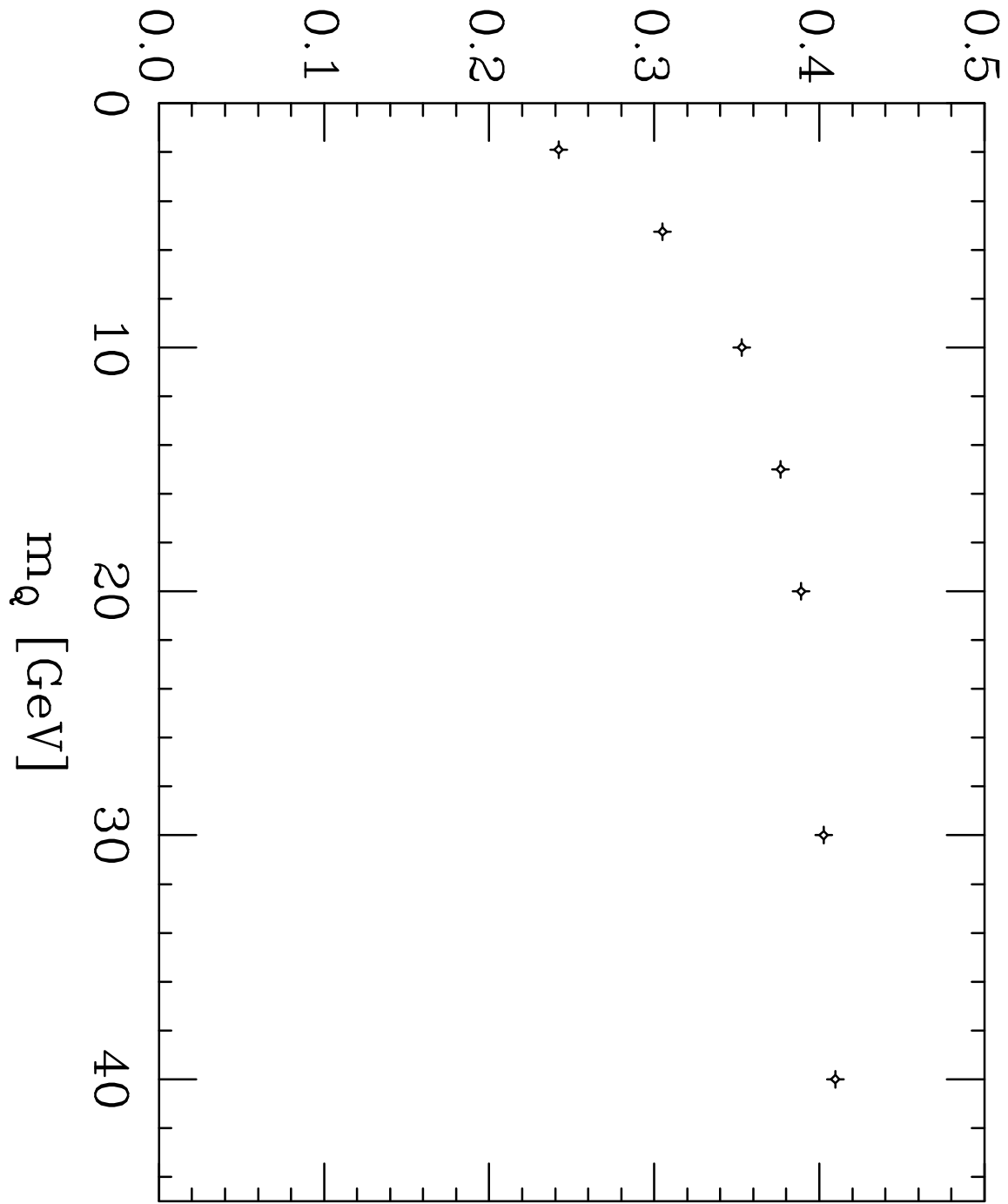
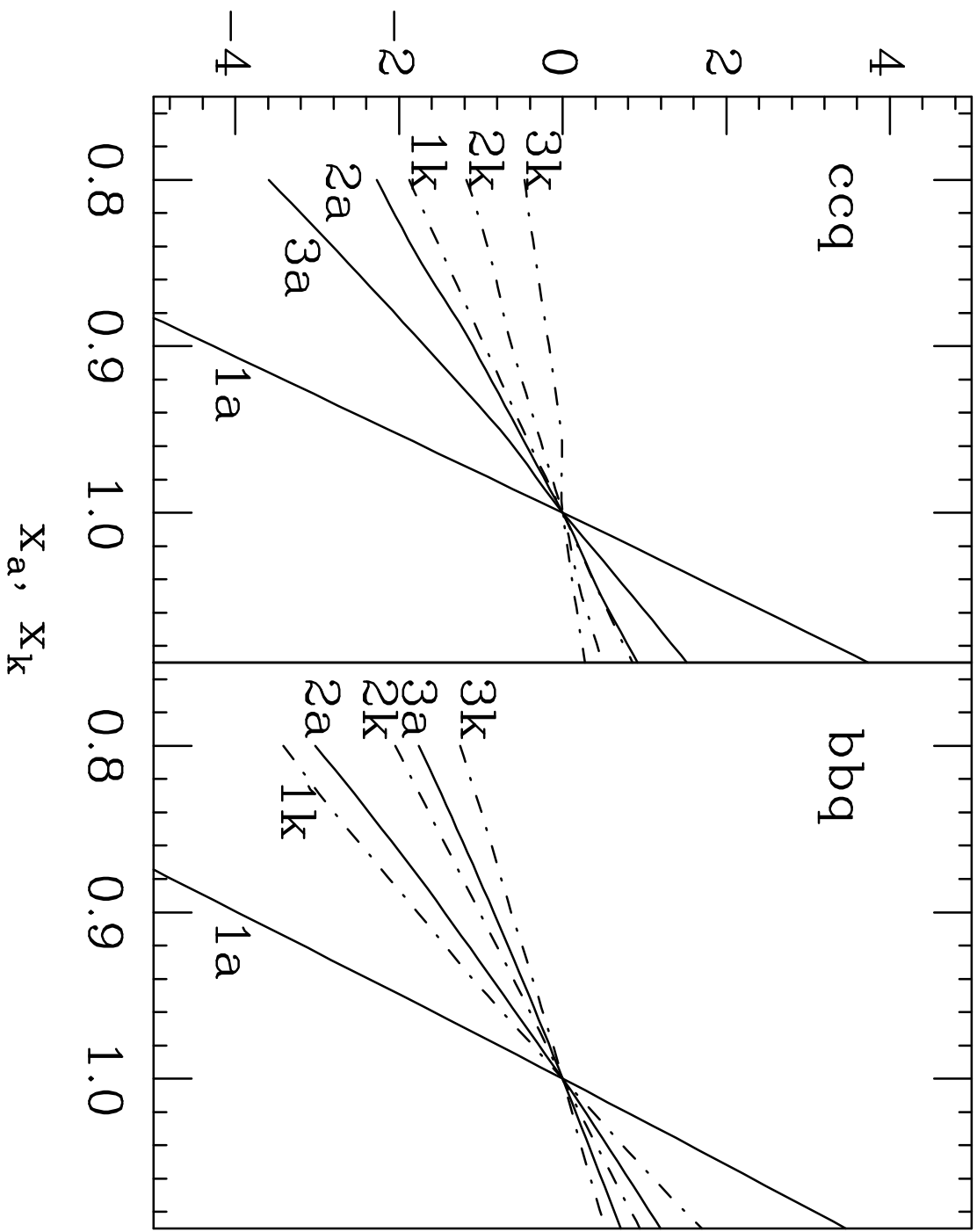
\end{document}